\documentclass[nopreprint,reprint,superscriptaddress,amsmath,amssymb,aps,prl]{revtex4-2}

\usepackage{graphicx}

\usepackage[dvipsnames,table,xcdraw]{xcolor}
\usepackage{tensor}
\usepackage[all,cmtip]{xy}
\usepackage{dcolumn}
\usepackage{bm}
\usepackage{tikz,tikz-cd}
\usepackage{quiver}

\newcommand{\overbar}[1]{\mkern 1.5mu\overline{\mkern-1.5mu#1\mkern-1.5mu}\mkern 1.5mu}
\def\TT{{ \mathrm{T} \overbar{\mathrm{T}}}}

\makeatletter
\newcommand*\bigcdot{\mathpalette\bigcdot@{.5}}
\newcommand*\bigcdot@[2]{\mathbin{\vcenter{\hbox{\scalebox{#2}{$\m@th#1\bullet$}}}}}
\makeatother

\newcommand{\deq}{\stackrel{\bigcdot}{=}}

\usepackage{tensor}
\usepackage{physics}
\usepackage{fontawesome}
\usepackage{slashed}
\usepackage{amsmath,amssymb,calc, amsthm,bbm, epsfig,psfrag, mathtools}
\usepackage[normalem]{ulem} 

\usepackage{latexsym,bm,amsfonts}
\usepackage{graphicx, enumerate}
 \usepackage[all,cmtip]{xy}
\allowdisplaybreaks

\newcommand{\Comment}[1]{{}}
\definecolor{darkblue}{rgb}{0.15,0.35,0.55}
\definecolor{reddish}{rgb}{0.65, 0.2, 0.2}
\usepackage[linktocpage=true]{hyperref}
\hypersetup{
colorlinks=true,
citecolor=darkblue,
linkcolor=reddish,
urlcolor=darkblue,
pdfauthor={},
pdftitle={},
pdfsubject={}
}

\def\be{\begin{equation}}
\def\ee{\end{equation}}
\def\bea{\begin{eqnarray}}
\def\eea{\end{eqnarray}}

\newfont{\goth}{ygoth.tfm scaled 1200}                   

\def\1{{(1)}}
\def\2{{(2)}}
\def\3{{(3)}}

\def\TT{{T\overbar{T}}}

\newcommand {\cT}{{\cal T}}

\def\g{\gamma}

\def\l{\lambda}

\newcommand{\ve}{\varepsilon}

\newcommand{\ba}{\begin{array}}
\newcommand{\ea}{\end{array}}

\def\double #1{#1{\hbox{\kern-2pt $#1$}}}

\newcommand{\bsubeq}{\begin{subequations}}
\newcommand{\esubeq}{\end{subequations}}

\def\tr{{\rm tr}}

\begin{document}

\title{
{\boldmath T-Duality and $\TT$-like Deformations of Sigma Models}}

\author{Daniele Bielli}
\email{d.bielli4@gmail.com}
\affiliation{High Energy Physics Research Unit, Faculty of Science, Chulalongkorn University, Bangkok 10330, Thailand}%

\author{Christian Ferko}\email{caferko@ucdavis.edu}
\affiliation{%
Center for Quantum Mathematics and Physics (QMAP),
Department of Physics \& Astronomy, University of California, Davis, CA 95616, USA.}

\author{Liam Smith}\email{liam.smith1@uq.net.au}
\affiliation{School of Mathematics and Physics, University of Queensland, St Lucia, Brisbane, Queensland 4072, Australia}%

\author{Gabriele Tartaglino-Mazzucchelli}
\email{g.tartaglino-mazzucchelli@uq.edu.au}
\affiliation{School of Mathematics and Physics, University of Queensland, St Lucia, Brisbane, Queensland 4072, Australia}%

\date{\today}

\begin{abstract}

We initiate the study of the interplay between T-duality and classical stress tensor deformations in two-dimensional sigma models. We first show that a general Abelian T-duality commutes with the $\TT$ deformation, which can be engineered by a gravitational dressing. Then, by using an auxiliary field formulation of stress tensor deformations of the principal chiral model (PCM), we prove that non-Abelian T-duality and arbitrary $\TT$-like flows also commute for theories in this class. We argue that all such auxiliary field deformations of both the PCM and its T-dual are classically integrable.

\end{abstract}

\maketitle

\section{Introduction}

Quantum field theory (QFT) is a common language used in many disciplines of modern theoretical physics, including high-energy particle physics, condensed matter theory, and statistical mechanics. Despite the numerous applications of this framework, our present understanding of the space of quantum field theories is fundamentally incomplete. On the one hand, sometimes two seemingly-different QFTs describe the same physical system; in this case, we say the two are related by a \emph{duality}. On the other hand, excluding isolated points in the space of QFTs, many quantum field theories occur in connected families whose properties are not fully understood. Such families are said to be related by \emph{deformations}

Two-dimensional sigma models offer a fertile playground for investigating these features, as QFTs in this class both enjoy dual descriptions and possess several interesting deformations. These models feature a long list of applications ranging from low- to high-energy physics, and in formal branches of mathematics such as geometry and topology. In the last few decades, sigma models have also played a key role in string theory, where \textit{target space} duality (T-duality) was first discovered in the context of toroidal compactifications \cite{Kikkawa:1984cp,Sakai:1985cs,Sathiapalan:1986zb}. Since then, T-duality has spurred the discoverey of deep connections in the web of string dualities, leading to new non-perturbative insights, and establishing further relations to mathematics via the discovery of mirror symmetry \cite{Hori:2000kt,Hori:2003ic}.
More recently, the non-Abelian extension of toroidal T-duality \cite{Fridling:1983ha,Fradkin:1984ai,delaOssa:1992vci,Giveon:1993ai,Alvarez:1993qi,Sfetsos:1994vz,Alvarez:1994np} has attracted a lot of attention, leading to manifestly duality-invariant formulations \cite{Klimcik:1995ux,Klimcik:1995jn,Hull:2009mi}, connecting with certain deformations of sigma models \cite{Hassan:1992gi,Henningson:1992rn,Klimcik:2002zj,Sfetsos:2013wia,Klimcik:2015gba,Osten:2016dvf,Hoare:2016wsk,Borsato:2016pas,Borsato:2017qsx,Borsato:2018idb,Osten:2019ayq,Borsato:2021gma}, allowing for novel examples in supergravity and holography \cite{Sfetsos:2010uq,Lozano:2011kb,Lozano:2012au,Itsios:2012zv,Kelekci:2014ima,Lozano:2014ata,Lozano:2015cra,Lozano:2015bra,Lozano:2016wrs,Lozano:2017ole,Ramirez:2021tkd,Ramirez:2022fkc} and leading to various supersymmetric generalisations in both the Abelian \cite{Berkovits:2008ic,Ricci:2007eq,Beisert:2008iq,Adam:2009kt,Hao:2009hw,Grassi:2009yj,Adam:2010hh,Bakhmatov:2010fp,Dekel:2011qw,Abbott:2015mla,Colgain:2016gdj} and non-Abelian sector \cite{Borsato:2016pas,Borsato:2017qsx,Borsato:2018idb,Grassi:2011zf,Eghbali:2009cp,Eghbali:2013bda,Eghbali:2023sak,Eghbali:2024vxi,Astrakhantsev:2021rhj,Astrakhantsev:2022mfs,Butter:2023nxm,Bielli:2021hzg,Bielli:2022gmm,Bielli:2023bnh,Bielli:2024wru}. See also \cite{Giveon:1994fu,OColgain:2012si,Bugden:2018pzv,Demulder:2019bha,Thompson:2019ipl,Seibold:2020ouf,Hoare:2021dix,Borsato:2023dis} for reviews on the above topics and further references.

Among deformations of QFTs, those that preserve the property of \emph{integrability} are of particular interest, since one can often solve for the dynamics of integrable field theories exactly (both at the classical and quantum levels).
In recent years, new integrable deformations based on composite operators constructed out of the stress-energy tensor
have opened a new avenue of research. For two-dimensional QFTs, the first example was the $T\overbar{T}$ deformation of  \cite{Zamolodchikov:2004ce, Smirnov:2016lqw,Cavaglia:2016oda}, but other stress tensor deformations such as root-$T\overbar{T}$, introduced in 
\cite{Rodriguez:2021tcz,Ferko:2022cix,Conti:2022egv,Babaei-Aghbolagh:2022leo,Babaei-Aghbolagh:2022uij}, have also shown remarkable features. Both $T\overbar{T}$ and root-$T\overbar{T}$  
preserve integrability, \cite{Smirnov:2016lqw,Borsato:2022tmu} (the latter at least at the classical level). $T\overbar{T}$ has also been shown to preserve supersymmetry \cite{Baggio:2018rpv,Chang:2018dge,Jiang:2019hux,Chang:2019kiu,Ferko:2021loo} and has played a new role in holography \cite{McGough:2016lol,Giveon:2017nie} (see also~\cite{Apolo:2019zai,Chakraborty:2020cgo,Coleman:2021nor,He:2024xbi}), a topic that has likewise been explored for root-$T\overbar{T}$ \cite{Ebert:2023tih,Ebert:2024zwv}.
Stress tensor deformations have also been studied in higher dimensions \cite{Taylor:2018xcy,Hartman:2018tkw,Conti:2018jho,Ferko:2019oyv,Babaei-Aghbolagh:2020kjg,Ferko:2023ruw,Ferko:2023sps,Ferko:2024zth,Babaei-Aghbolagh:2024uqp}.

A streamlined method for engineering $T\overbar{T}$ in arbitrary dimensions makes use of an auxiliary metric (vielbein) \cite{Coleman:2019dvf,Tolley:2019nmm,Conti:2022egv,Morone:2024ffm,Babaei-Aghbolagh:2024hti,Tsolakidis:2024wut,Ferko:2024yhc,Brizio:2024arr}. This method, which is referred to as gravitational dressing, remarkably connects the study of deformations of quantum field theories to $2d$ (topological) gravity \cite{Dubovsky:2017cnj,Cardy:2018sdv,Dubovsky:2018bmo,Conti:2018tca,Conti:2019dxg,Caputa:2020lpa,Conti:2022egv,Bhattacharyya:2023gvg}, and the broad literature on $d$-dimensional models of massive gravity \cite{Tolley:2019nmm,Morone:2024ffm,Babaei-Aghbolagh:2024hti,Tsolakidis:2024wut}.

The use of auxiliary fields is ubiquitous in field theory. Some examples include off-shell supersymmetry \cite{Gates:1983nr,Wess:1992cp,Buchbinder:1998qv}, chiral (tensor) fields in various dimensions \cite{Siegel:1983es,Pasti:1995tn,Pasti:1996vs,Pasti:1997gx,Ivanov:2014nya,Buratti:2019cbm,Buratti:2019guq,Sen:2015nph,Sen:2019qit,Townsend:2019koy,Mkrtchyan:2019opf,Bansal:2021bis,Avetisyan:2021heg,Avetisyan:2022zza,Evnin:2022kqn,Arvanitakis:2022bnr,Ferko:2024zth}, and theories of $4d$ duality-invariant non-linear electrodynamics \cite{Ivanov:2002ab,Ivanov:2003uj,Ferko:2023wyi}. Earlier in 2024, integrable stress tensor deformations of the $2d$ principal chiral model (PCM) were engineered through new auxiliary field sigma models (AFSM) \cite{Ferko:2024ali} -- see also \cite{Fukushima:2024nxm,Bielli:2024ach} for recent extensions. The PCM is a very natural playground for exploring aspects of non-Abelian T-duality, as well as various physically interesting extensions such as WZW models, (semi-)symmetric spaces, and so forth.

For sigma models, T-duality can be implemented through a gauging procedure of a subgroup H$\subseteq$G of the group G of target space isometries \cite{Rocek:1991ps,delaOssa:1992vci} (see also \cite{Alvarez:1994wj,Curtright:1994be,Lozano:1995jx}).
Considering that stress tensor deformations are expected to preserve symmetries -- albeit perhaps in a modified or ``dressed'' form, as occurs, e.g., for conformal symmetry 
\enlargethispage*{3\baselineskip}
\cite{Guica:2020uhm,Guica:2021pzy,Guica:2022gts} -- it is natural to argue that T-duality and $T\overbar{T}$-like deformations should ``commute''. By using the auxiliary field formulations mentioned above, in our letter, we explicitly show how this expectation is realized. As a byproduct of part of our analysis, we show how stress tensor deformations of the PCM preserve integrability irrespective of the order in which T-duality and deformations are performed. Ultimately, our analysis and results have the scope of initiating the systematic study of T-duality and stress tensor deformations.

\section{Abelian T-duality and $\TT$ Gravitational dressing}
\label{secII}

In this section, we discuss Abelian T-duality in $\TT$-deformed sigma models by using a gravitational dressing.

Consider a sigma model where the fields $\phi^I$, with $I=1,\cdots,D$, are a set of local coordinates for a target space of dimension $D$. The $2d$ sigma model action is
\bea\label{undeformed_sigma}
S^{(0)}_{\rm sm}
 =
 -\frac{1}{2}\int \mathrm{d}^2x\,
\Big\{ e\,g^{\mu\nu}\Phi_{\mu\nu}
+\mathcal{T}\Big\}\,, 
\label{sigmaM-1}
\eea
where 
\bea
\Phi_{\mu\nu}
:=
G_{IJ} \partial_{\mu}\phi^{I} \partial_{\nu}\phi^{J}
\,,~~~
\cT
:=
\ve^{\mu\nu}
B_{IJ} \partial_{\mu}\phi^{I} \partial_{\nu}\phi^{J}
\,.
\eea
Here, $G_{IJ}=G_{JI}$ is the target space metric, $B_{IJ}=-B_{JI}$ is the Kalb-Ramond field, $e_\mu{}^a$ is the zweibein of a worldsheet with metric $g_{\mu\nu}=e_{\mu}{}^a e_{\nu}{}^b\eta_{ab}$, and $\cT$ is a $2d$ topological term. For simplicity, we do not consider a potential for the scalar fields in \eqref{sigmaM-1}, though this would not change the key features of the analysis in this section.

For the sigma model \eqref{sigmaM-1}, T-duality can be implemented by gauging and dualising an Abelian isometry of the target space \cite{Buscher:1987sk,Buscher:1987qj,Rocek:1991ps,delaOssa:1992vci}. We choose the direction of the Abelian isometry to be parametrised by a local coordinate $\varphi$ in a coordinate system where the isometry is a shift symmetry. Then we have the splitting of indices $I=(\hat{I},\varphi),~\hat{I}=1,\cdots,D-1$.
Without repeating the derivation of \cite{Buscher:1987sk,Buscher:1987qj}, 
we simply state the resulting Buscher rules for Abelian T-duality. These are given by 
\bsubeq\label{Buscher}
\bea
&\tilde{G}_{\varphi\varphi}
=
G^{-1}_{\varphi\varphi}
\,,~~
\tilde{G}_{\hat{i}\varphi}
=
G^{-1}_{\varphi\varphi}
B_{\hat{i}\varphi}
\,,~~
\tilde{B}_{\hat{i}\varphi}
=
G^{-1}_{\varphi\varphi}
G_{\hat{i}\varphi}
\,,~~~~~~
\\
&\tilde{G}_{\hat{i}\hat{j}}
=
G_{\hat{i}\hat{j}}
-
G^{-1}_{\varphi\varphi}(G_{\varphi\hat{i}}G_{\varphi\hat{j}}-B_{\varphi\hat{i}}B_{\varphi\hat{j}})
\,,
\\
&\tilde{B}_{\hat{i}\hat{j}}
=
G_{\hat{i}\hat{j}}
-
G^{-1}_{\varphi\varphi}(
G_{\varphi\hat{i}}G_{\varphi\hat{j}}-G_{\varphi\hat{i}}B_{\varphi\hat{j}})
\,,
\eea
\esubeq
where $\tilde{G}_{IJ}$ and $\tilde{B}_{IJ}$ are the metric and Kalb-Ramond fields after T-duality. It is important to stress that $e_\mu{}^a$ is unaffected by T-duality and the dual sigma model is
\begin{equation}
\tilde{S}^{(0)}_{\rm sm}
 =
 -\frac{1}{2}\int \mathrm{d}^2x\,
 \big{(}e g^{\mu\nu}\tilde{G}_{IJ}
 +\ve^{\mu\nu}\tilde{B}_{IJ}\big{)}
 \partial_{\mu}\phi^{I} \partial_{\nu}\phi^{J}
\,.
\end{equation}
If we were to consider a consistent quantum sigma model, it would also be necessary to include a dilaton coupling to cancel tadpoles. The dilaton would also transform under T-duality \cite{Buscher:1987sk,Buscher:1987qj}. However, our letter concerns only a classical analysis, and we ignore the dilaton.

Let us now turn our attention to $\TT$. As mentioned in the introduction, gravitational dressing refers to engineering $\TT$ through a zweibein auxiliary field. By following \cite{Tolley:2019nmm,Babaei-Aghbolagh:2024hti,Tsolakidis:2024wut}, given a generic matter model with action $S^{(0)}[\phi,e_\mu{}^a]$, the $\TT$ deformed theory is 
\bea
S^{(\l)}[\phi,e_\mu{}^a,f_\mu{}^a]=S^{(0)}[\phi,e_\mu{}^a]+S_{G}[e_\mu{}^a,f_\mu{}^a,\lambda]
\,,
\label{Szwei-0}
\eea
with
\bea
S_{G}=\int \mathrm{d}^2x\,\frac{1}{2\l}\ve^{\mu\nu}\ve_{ab}(e_\mu{}^a-f_\mu{}^a)(e_\nu{}^b-f_\nu{}^b)
~.
\label{Sgrav-0}
\eea
Here $e_\mu{}^a$ is an auxiliary field while $f_\mu{}^a$ is the zweibein of a background geometry, which in our case will always be $2d$ Minkowski space-time. Once $e_\mu{}^a$ is integrated out, one obtains the deformed theory $S^{(\l)}[\phi,f_\mu{}^a]=S^{(\l)}[\phi,f_\mu{}^a,e^*_\mu{}^a]$, with $e^*_\mu{}^a=e_\mu{}^a(\phi,f_\nu{}^b)$ being the solution of the $e_\mu{}^a$ equation of motion derived from \eqref{Szwei-0}. The action $S^{(\l)}[\phi,f_\mu{}^a]$ satisfies
\bea
\frac{\mathrm{d} S^{(\l)}[\phi,f_\mu{}^a]}{\mathrm{d}\lambda}
=-\int \mathrm{d}^2x\,f\,\det[T^\mu{}_{\nu}]
\,,
\eea
where $T^\mu{}_{\nu}$ is the Hilbert stress-energy tensor defined through the variation with respect to $f_\mu{}^a$ \cite{Babaei-Aghbolagh:2024hti,Tsolakidis:2024wut}.

By applying the gravitational dressing, the $\TT$ deformed action for the sigma model \eqref{sigmaM-1} is given by \eqref{Sgrav-0} with $S_{\rm sm}^{(\lambda)}=S^{(0)}_{\rm sm}+S_{G}$. In this case, it is possible to explicitly solve the EOM of $e_m{}^a$. The solution is \cite{Tsolakidis:2024wut}
\begin{gather}
        \tensor{{({e}^*_{\pm})}}{_{\mu}^{a}}=\frac{1}{2}\tensor{{f}}{_{\mu}^{a}}
 \pm
 \frac{1}{2}\frac{\tensor{{f}}{_{\mu}^{a}}+2{\lambda}\tensor{\Phi}{_{\mu}_{\nu}}\tensor{{f}}{^{\nu}_{b}}\tensor{\eta}{^{b}^{a}}}{\sqrt{\g^{-1}\det(\tensor{{\g}}{_{\mu}_{\nu}}+2{\lambda}\tensor{\Phi}{_{\mu}_{\nu}})}}\,, 
 \label{grav-dressing-1}
\end{gather}
where $\tensor{{\g}}{_{\mu}_{\nu}}\coloneqq\tensor{{f}}{_{\mu}^{a}}\tensor{{f}}{_{\nu}^{b}}\tensor{\eta}{_{a}_{b}}$. Note that the $B$-field does not appear in \eqref{grav-dressing-1} since $\cT$ is a topological term. Therefore, $\cT$ is not affected by the $\TT$ deformation.

At this stage, we have two operations that can be performed. One is T-duality while the other is a $\TT$ deformation. A natural question is: do these two operations commute? In the gravitational dressing, it is straightforward to see that they do. In fact, the action $S_{G}[e_\mu{}^a,f_\mu{}^a,\l]$, and the zweibein $e_\mu{}^a$ in $S^{(\l)}_{\rm sm}$ are inert under T-duality while $S^{(0)}_{\rm sm}$ simply turns into $\tilde{S}^{(0)}_{\rm sm}$ by using the undeformed Buscher rules \eqref{Buscher}. Remarkably, all complications of the $\TT$ deformation are hidden in $S_{G}$ and the integration of $e_\mu{}^a$, which leads to a dressing of the sigma model with an involved dependence of higher-derivative terms. From this point of view, the target space geometry and its T-dual are unaffected by $\TT$.

Importantly, note that the previous commutativity argument works for any $\TT$-like deformation engineered by gravitational dressing through an action $S_{G}[e_\mu{}^a,f_\mu{}^a,\lambda]$ and equation \eqref{Szwei-0}, not just $\TT$. By changing $S_{G}$, the deformation would also change the explicit solution \eqref{grav-dressing-1}; see \cite{Babaei-Aghbolagh:2024hti,Tsolakidis:2024wut} for examples, including root-$\TT$. 
There is, however, a limitation. Though the arguments in this section could be generalised to non-Abelian T-duality, 
\enlargethispage*{3\baselineskip}
this would work only for $\TT$-like deformations that admit a gravitational dressing $S_{\rm sm}^{(\lambda)}=S^{(0)}_{\rm sm}+S_{G}$ for some $S_{G}$ independent of the matter fields $\phi^I$. By choosing an alternative auxiliary field approach to $\TT$-like deformations, we will easily overcome this difficulty, at least for a subclass of interesting sigma models.

\section{Non-Abelian T-duality and Auxiliary Field Sigma Models}
\label{secIII}

In this section we extend the T-dualisation procedure of $T\overline{T}$ deformed models to the non-Abelian setting, exploiting the infinite family of integrable deformations introduced in \cite{Ferko:2024ali}. We start by briefly reviewing the AFSM construction. These are $2d$ sigma models on a flat Lorentzian worldsheet $\Sigma$, whose target-space geometry is a Lie group G with Lie algebra $\mathfrak{g}$, which are characterized by the pullback to $\Sigma$ of the left-invariant Maurer-Cartan form $j:= \texttt{g}^{-1}\mathrm{d}\texttt{g}$. The standard PCM action is
\begin{equation}\label{PCM-action}
S_{\text{PCM}} \!:=\!  \int_{\! \Sigma\!} \! \mathrm{d}^2 \! \sigma  \mathcal{L}_{\text{PCM}}
\,\,\,\text{with} \,\,\,\,
\mathcal{L}_{\text{PCM}} \!:= - \frac{1}{2} \mathrm{tr}(j_{+}j_{-})
\,.
\end{equation}
Here $\sigma^{\pm}:= \tfrac{1}{2}(\tau \pm \sigma)$ are worldsheet lightcone coordinates, in terms of which the metric and its inverse reduce to $\eta_{+-}=\eta_{-+}=-2$ and $\eta^{+-}=\eta^{-+}=-\tfrac{1}{2}$ while $\mathrm{tr}$ is the trace of the Lie algebra generators in some representation. The infinite family of deformed models is described by the Lagrangian
\begin{equation}\label{deformed-PCM}
\mathcal{L}_{\text{PCM}}^{\text{E}} 
:=  
\mathrm{tr}\Big{[}
\tfrac{1}{2}j_{+}j_{-}
\!+\! v_{+}v_{-}
\!+\!j_{+}v_{-}+j_{-}v_{+}\Big{]}
\!+\!E(\nu)
\,,
\end{equation}
where $v$ is the auxiliary field, a Lie-algebra-valued 1-form, and $E$ is an arbitrary function of $\nu:= \mathrm{tr}(v_{+}v_{+})\mathrm{tr}(v_{-}v_{-})$.
The equations of motion (EOM) of the model read
\begin{align}
\delta_{\texttt{g}}\mathcal{L}_{\text{PCM}}^{\text{E}} \!\equiv\! 0 
~&\leftrightarrow~
\partial_{\!+\!}\mathfrak{J}_{\!-\!}\!+\!\partial_{\!-\!}\mathfrak{J}_{\!+\!}\! =\!-2([v_{\!+\!},j_{\!-\!}]\!+\![v_{\!-\!},j_{\!+\!}]) \, ,
\notag \\
\delta_{v_{\!\pm\!}}\mathcal{L}_{\text{PCM}}^{\text{E}} \!\equiv\! 0 
~&\leftrightarrow ~
j_{\pm}\! =\!-v_{\pm}\!-\!2E'v_{\mp}\mathrm{tr}(v_{\pm}v_{\pm})
\,,
\label{auxiliary-fields-EOM-initial-model}
\end{align}
with $E':= \tfrac{\mathrm{d}E}{\mathrm{d}\nu}$ and $\mathfrak{J}_{\pm}:=-(j_{\pm}+2v_{\pm})$. For $E \!=\! 0$ one correctly recovers \eqref{PCM-action} after solving the second equation in \eqref{auxiliary-fields-EOM-initial-model} for $v_{\pm}$. 
In this language, the non-linearities of the deformation are encoded in the arbitrary interaction function $E$. Analogously to the gravitational dressing of the previous section, the T-dualisation introduced in
\cite{Rocek:1991ps,delaOssa:1992vci} as a generalisation of \cite{Buscher:1987sk,Buscher:1987qj},  becomes as simple as in the undeformed case. The description in terms of the left-invariant Maurer-Cartan form ensures preservation of the standard G$_{L}\times$G$_{R}$ isometry group of the undeformed PCM ($\texttt{g}\rightarrow g_{L}^{-1}\texttt{g}g_{R}$, $j \rightarrow g_{R}^{-1}jg_{R}$) provided the auxiliary field transform as $v\rightarrow g_{R}^{-1}vg_{R}$. It is then straightforward to gauge a subgroup H$\subseteq$G$_{L}$ of the left sector of the isometry group via a minimal coupling 
\begin{equation}
j \quad \rightarrow \quad j^{\omega} := \texttt{g}^{-1}(\mathrm{d}+\omega) \texttt{g} = j + \texttt{g}^{-1}\omega \texttt{g} \,.
\end{equation}
The connection $\omega$ is invariant under global G$_{R}$ transformations, while $\omega \!\rightarrow\! h^{-1}\!\omega h \!+\! h^{-1}\!\mathrm{d} h$ under local left action of $h \in{\rm H}\subseteq{\rm G}_{L}$. The gauge invariant Lagrangian is
\begin{equation}\label{gauged-action}
\mathcal{L}_{\omega} \!=\! 
\mathrm{tr}\Big{[}
\tfrac{1}{2}j^{\omega}_{+}j^{\omega}_{-}
\!+\! v_{+}v_{-} 
\!+\! j^{\omega}_{+}v_{-} 
\!+\! j^{\omega}_{-}v_{+}
\Big{]}
\!+\! E(\nu)\,.
\end{equation}
Proceeding with the dualisation requires adding Lagrange multipliers which enforce the flatness of $\omega$:
\begin{equation}\label{multipliers-term}
\mathcal{L}_{\Lambda} = \tfrac{1}{2}\mathrm{tr}(\Lambda F_{\omega}) = \tfrac{1}{2}\mathrm{tr}(X F_{j_\omega}) \,\,\, \text{with} \,\,\, X:=\texttt{g}^{-1}\Lambda \texttt{g} \,.
\end{equation}
In the previous equation  we exploited the relation
\begin{equation}
F_{j_\omega}=F_{j}+\texttt{g}^{-1}F_{\omega}\texttt{g} \,\,\,\,\, \text{with} \,\,\,\,\, F_{A}:=\mathrm{d}A+\tfrac{1}{2}[A,A] \,,
\end{equation}
and $F_{j}=0$ by construction. For $\mathcal{L}_{\Lambda}$ to respect the symmetries of $\mathcal{L}_{\omega}$, the multipliers should transform as $\Lambda \rightarrow h^{-1}\Lambda h$ under local H and global G$_R$ transformations. Notice that while $\Lambda \in \mathfrak{h}$, the above identity leads to $X \in \mathfrak{g}$, so that depending on which subgroup H$\subseteq$G$_{L}$ has been gauged, $X$ generically contains, even after gauge fixing, a mixture of pure Lagrange multipliers $\Lambda$ and original coordinates inherited from $\texttt{g}$. Upon gauging H$=$G$_{L}$ one has the freedom to completely eliminate the initial coordinates by setting $\texttt{g}=\mathrm{1}$, such that $X\equiv \Lambda$. 

In lightcone coordinates, the multiplier term \eqref{multipliers-term} reads
\begin{equation}\label{multiplier-term-lightcone}
\mathcal{L}_{\Lambda}=\tfrac{1}{2}\mathrm{tr}\Bigl[ X(\partial_{+}j^{\omega}_{-}-\partial_{-}j^{\omega}_{+}+[j^{\omega}_{+},j^{\omega}_{-}])\Bigr]
\,,
\end{equation}
and the total Lagrangian, also known as \textit{Master Lagrangian}, is the sum of \eqref{gauged-action} and \eqref{multiplier-term-lightcone}, $\mathcal{L}_{\mathrm{Master}}=\mathcal{L}_{\omega}+\mathcal{L}_{\Lambda} $. To proceed toward the T-dual model, we perform integration by parts on the derivative terms in \eqref{multiplier-term-lightcone} and then integrate out the gauge fields. Their EOM take the form
\begin{equation}\label{gauge-fields-EOM}
(1\pm\mathrm{ad}_{X})j^{\omega}_{\pm}=\pm (\partial_{\pm}X\mp 2v_{\pm})\,,
\end{equation}
and can readily be solved as 
\begin{equation}\label{gauge-fields-EOM-solution}
j^{\omega}_{\pm} = \pm \frac{1}{1 \pm  \mathrm{ad}_{X}}(\partial_{\pm}X \mp 2v_{\pm}) \,.
\end{equation}
Substituting \eqref{gauge-fields-EOM-solution} back into $\mathcal{L}_{\text{Master}}$, and rearranging terms, one obtains the T-dual model
\begin{equation}\label{T-dual-action}
\begin{aligned}
\tilde{\mathcal{L}}&=\tfrac{1}{2}\mathrm{tr}\Bigl[ (\partial_{+}X-2v_{+})\frac{1}{1-\mathrm{ad}_{X}}(\partial_{-}X+2v_{-}) \Bigr]
\\
& \,\,\,\,\, + \mathrm{tr}(v_{+}v_{-})+E(\nu) \,.
\end{aligned}
\end{equation}

\section{Deformation of T-dual models}
\label{secIV}

In the last section, we have obtained \eqref{T-dual-action} by first deforming the PCM action \eqref{PCM-action} as in \eqref{deformed-PCM}, and successively T-dualising. Do these operations commute, as with the gravitational dressing?
We will 
see here that the Abelian results still hold, up to implementing a field redefinition of the vector auxiliary fields in \eqref{T-dual-action}.

We start by defining 
\begin{equation}
v_{\pm} := \mp \frac{1}{1\mp\mathrm{ad}_{X}} (\tilde{v}_{\pm}) \,.
\end{equation}
It is not hard to realise that \eqref{T-dual-action} becomes
\begin{align}\label{field-redefined-dual-lagrangian}
\tilde{\mathcal{L}}&=\tfrac{1}{2}\mathrm{tr} \Bigl( \partial_{+}X\frac{1}{1-\mathrm{ad}_{X}^2}\partial_{-}X \Bigr)+\tfrac{1}{2}\mathrm{tr} \Bigl( \partial_{+}X\frac{\mathrm{ad}_{X}}{1-\mathrm{ad}_{X}^2}\partial_{-}X \Bigr)
\notag \\
& \,\,\,\, + \mathrm{tr} \Bigl( \tilde{v}_{+} \frac{1}{1-\mathrm{ad}_{X}^2}\partial_{-}X \Bigr) + \mathrm{tr} \Bigl( \partial_{+}X\frac{1}{1-\mathrm{ad}_{X}^2}\tilde{v}_{-} \Bigr)
\notag \\
& \,\,\,\, + \mathrm{tr} \Bigl( \tilde{v}_{+}\frac{1}{1-\mathrm{ad}_{X}^2}\tilde{v}_{-} \Bigr) + E(\tilde{\nu}) \,,
\end{align}
with 
\begin{equation}\label{mu-definition}
\tilde{\nu}:= \mathrm{tr} \Bigl( \tilde{v}_{+}\frac{1}{1-\mathrm{ad}_{X}^2}\tilde{v}_{+} \Bigr)\mathrm{tr} \Bigl( \tilde{v}_{-}\frac{1}{1-\mathrm{ad}_{X}^2}\tilde{v}_{-} \Bigr) \,.
\end{equation}
The Lagrangian \eqref{field-redefined-dual-lagrangian} has then the precise structure of a general auxiliary field deformation of
\begin{equation}\label{undeformed-PCM-T-dual}
\tilde{\mathcal{L}} \!=  -  \tfrac{1}{2}\mathrm{tr} \Bigl[ \partial_{-}X  \frac{1}{1 \!-\! \mathrm{ad}_{X}^2 \!} \partial_{+}X
+\partial_{-}X \frac{\mathrm{ad}_{X}}{ 1 \!-\! \mathrm{ad}_{X}^2 \!} \partial_{+}X  \Bigr]
\,,
\end{equation}
which corresponds to the Lagrangian obtained by T-dualising the standard PCM \eqref{PCM-action}, upon splitting $\tfrac{1}{1-\mathrm{ad}_X}$ into its symmetric and antisymmetric components. One can indeed first check that, for $E=0$, the EOM of the auxiliary field $\tilde{v}_{\pm}$ in \eqref{field-redefined-dual-lagrangian} and $v_{\pm}$ in \eqref{T-dual-action} both lead to the undeformed T-dual Lagrangian \eqref{undeformed-PCM-T-dual}. Secondly, it is not hard to find that the stress tensor of \eqref{undeformed-PCM-T-dual} satisfies
\begin{equation}
T^{\!\alpha\beta\!}T_{\!\alpha\beta\!} \! = \! \mathrm{tr} \Bigl( \! \partial_{+}X \frac{1}{\! 1 \!-\! \mathrm{ad}_{X}^2 \!}\partial_{+}X \! \Bigr)\mathrm{tr} \Bigl( \! \partial_{-}X\frac{1}{\! 1\!-\! \mathrm{ad}_{X}^2 \!} \partial_{-}X \! \Bigr) 
\end{equation}
corresponding, as expected, to the structure observed in the variable $\tilde{\nu}$ in \eqref{mu-definition}, on which the function $E$ depends in \eqref{field-redefined-dual-lagrangian}. Finally, one can compute the stress tensor for \eqref{field-redefined-dual-lagrangian} and construct the two independent invariants
\begin{equation}
\begin{aligned}
T_{\alpha}{}^{\alpha}&\!=\!2(E\!-\!2\tilde{\nu} E')\,,
\\
T^{\!\alpha\beta\!}T_{\!\alpha\beta\!} & \!=\! 2(E\!-\!2\tilde{\nu} E')+\frac{\tilde{\nu}}{2}\Bigl( 1-4\tilde{\nu}(E')^2 \Bigr)\,.
\end{aligned}
\end{equation}
These are the same as the ones found in \cite{Ferko:2024ali} for \eqref{deformed-PCM} and lead to the same ordinary differential equations for the interaction functions. This confirms the commutativity up to auxiliary fields redefinitions, as illustrated in the diagram below.

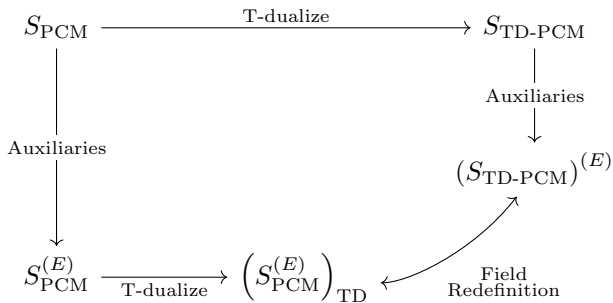
\begin{align*}
\begin{tikzcd}[ampersand replacement=\&]
        {S_{\text{PCM}}} \&\&\& {S_{\text{TD-PCM}}} \\
        \\
        \&\&\& {\left( S_{\text{TD-PCM}} \right)^{(E)}} \\
        {S^{(E)}_{\text{PCM}}} \&\& {\left( S^{(E)}_{\text{PCM}} \right)_{\text{TD}}}
        \arrow["{\text{T-dualize}}", from=1-1, to=1-4]
        \arrow["{\text{Auxiliaries}}"{description}, from=1-1, to=4-1]
        \arrow["{\text{Auxiliaries}}"{description}, from=1-4, to=3-4]
        \arrow["{\text{T-dualize}}"', from=4-1, to=4-3]
        \arrow["{\substack{\large \text{Field} \\ \normalsize \text{Redefinition}}}"', curve={height=18pt}, tail reversed, from=4-3, to=3-4]
\end{tikzcd}
\end{align*}

\section{Integrability}
\label{secV}

In this section, we discuss the integrable structure underlying \eqref{deformed-PCM} and \eqref{T-dual-action}. We start by recalling results and notation from \cite{Ferko:2024ali}.
For the model \eqref{deformed-PCM} with any $E(\nu)$,
the first EOM in \eqref{auxiliary-fields-EOM-initial-model} represents a true dynamical condition on $\texttt{g}$, while the second one is a constraint imposed by the auxiliary field. The symbol $ \deq $ is used to denote equality upon satisfaction of the latter condition. For the inital model \eqref{deformed-PCM} one has $[v_{-},j_{+}] \,  \deq \, -[v_{+},j_{-}]$, which implies that the EOM for $\texttt{g}$ is not the conservation of $j$, as for the PCM, but rather $\mathfrak{J}:=-(j+2v)$ is conserved. Conversely, the flatness of $j$ is unaffected by $v$. The auxiliary field EOM also implies that $[\mathfrak{J}_{+},j_{-}] \, \deq \, [j_{+},\mathfrak{J}_{-}]$ and $[\mathfrak{J}_{+},\mathfrak{J}_{-}] \, \deq \, [j_{+},j_{-}]$, which, in turn, imply that flatness of $j$ and conservation of $\mathfrak{J}$ arise from the flatness $\partial_{+}\mathfrak{L}_{-}-\partial_{-}\mathfrak{L}_{+}+[\mathfrak{L}_{+},\mathfrak{L}_{-}] =0$
of the Lax connection
\begin{equation}
\mathfrak{L}_{\pm}:= \frac{j_{\pm}\pm z\mathfrak{J}_{\pm}}{1-z^2} \,.
\label{Lax-i}
\end{equation}
The Lax connection ensures the existence of an infinite set of conserved quantities. For the models \eqref{deformed-PCM}, one can further prove that the charges are Poisson-commuting by showing that the Poisson bracket $\{ \mathfrak{L}_{\sigma,1}(\sigma,z),\mathfrak{L}_{\sigma,2}(\sigma',z') \}$ takes the non-ultralocal Maillet form \cite{Ferko:2024ali}.

We now show how T-duality preserves the integrable structure of \eqref{deformed-PCM} by exchanging the role of the EOM and Maurer-Cartan equation in the T-dual model \eqref{T-dual-action} 
(see also \cite{Ivanov:1987yv}, where this interchange is interpreted as giving a dual description related to the Drinfeld double).
We start by computing the EOM for the T-dual fundamental field $X$, which take the form of a flatness condition 
\begin{equation}\label{T-dual-EOM-X}
\delta_{X}\tilde{\mathcal{L}}\equiv 0 \quad \leftrightarrow \quad \partial_{+}\tilde{j}_{-}-\partial_{-}\tilde{j}_{+}+[\tilde{j}_{+},\tilde{j}_{-}]=0 \,,
\end{equation}
after defining $\tilde{j}_{\pm}:= j^{\omega}_{\pm}$, given in \eqref{gauge-fields-EOM}. It follows that, as in the undeformed setting, $\tilde{j}$ satisfies by construction the gauge field EOM \eqref{gauge-fields-EOM}, which can be used to study its conservation. Indeed, from \eqref{gauge-fields-EOM} one obtains
\begin{align}
\partial_{\pm} \tilde{j}_{\mp} 
& = [(1\pm \mathrm{ad}_{X})\tilde{j}_{\pm},\tilde{j}_{\mp}] + 2[v_{\pm},\tilde{j}_{\mp}]
\notag \\
& \,\,\,\,\, \pm [X,\partial_{\pm}\tilde{j}_{\mp}] \mp \partial_{+}\partial_{-} X - 2\partial_{\pm}v_{\mp} \, ,
\end{align}
and summing up the two contributions above leads to 
\begin{align}\label{conservation-dual-current}
& \partial_{+} \tilde{j}_{-} \!+\! \partial_{-} \tilde{j}_{+}
\! =\! -2 \Bigl( \partial_{+}v_{-} \!+\! \partial_{-}v_{+} \!+\! [\tilde{j}_{+},v_{-}] \!+\! [\tilde{j}_{-},v_{+}] \Bigr)
\notag
\\
& \,\, 
+ \bigl[\tilde{j}_{+},[X,\tilde{j}_{-}] \bigr]
\!-\!\bigl[\tilde{j}_{-},[X,\tilde{j}_{+}] \bigr] 
\!+\! \bigl[ X, \partial_{+}\tilde{j}_{-} 
\!\!-\! \partial_{-}\tilde{j}_{+}\bigr] \, .
\end{align}
Using the T-dual EOM \eqref{T-dual-EOM-X} and Jacobi identity for $(X,\tilde{j}_{+},\tilde{j}_{-})$, equation \eqref{conservation-dual-current} becomes
the conservation 
\begin{equation}
\partial_{+} \tilde{\mathfrak{J}}_{-} + \partial_{-} \tilde{\mathfrak{J}}_{+} = 0 \quad \text{with} \quad \tilde{\mathfrak{J}}_{\pm} := - (\tilde{j}_{\pm}+2v_{\pm}) \, .
\end{equation}
This result shows that the exchange in the role of EOM and Maurer-Cartan equation, which takes place in the undeformed case, is slightly modified in the presence of the auxiliary field. Though it might seem surprising, this is, in fact, perfectly in line with the analysis of \cite{Ferko:2024ali} for the deformed initial model \eqref{deformed-PCM}. There, it was shown that $v$ prevents $j$ from being conserved on-shell and forces the definition of a new conserved quantity $\mathfrak{J} := - (j+2v)$.

One may now hope to proceed in constructing the T-dual Lax connection by simply replacing $j$ with $\tilde{j}$ in the Lax connection \eqref{Lax-i}. While in the undeformed case this procedure automatically ensures that flatness and conservation of $\tilde{j}$ follow from the flatness of the dual Lax connection, in the deformed models one needs to pay a little extra care. In the initial deformed model, the flatness of the Lax connection implies flatness of $j$ and conservation of $\mathfrak{J}$ only after making use of the auxiliary field EOM. For this reason, one should expect that the T-dual Lax connection obtained by replacing $j$ with $\tilde{j}$ in \eqref{Lax-i} should imply flatness of $\tilde{j}$ and conservation of $\tilde{\mathfrak{J}}$ only after using the EOM for $v$. This expectation is readily verified by noting that the auxiliary field EOM take the exact same structure as in the initial model, namely
\begin{equation}
\delta_{v_{\pm}}\tilde{\mathcal{L}}\equiv 0 ~\iff ~\tilde{j}_{\pm} + v_{\pm} +2E'(\nu)\mathrm{tr}(v_{\pm}v_{\pm})v_{\mp}=0 \, .
\end{equation}
This allows us to verify the relations
$[v_{-},\tilde{j}_{+}]\deq -[v_{+},\tilde{j}_{-}]$,
$[\tilde{\mathfrak{J}}_{+},\tilde{j}_{-}]\deq [\tilde{j}_{+},\tilde{\mathfrak{J}}_{-}]$, and
$[\tilde{\mathfrak{J}}_{+},\tilde{\mathfrak{J}}_{-}]   \deq [\tilde{j}_{+},\tilde{j}_{-}]$ in the T-dual model, and to show that the Lax connection
\begin{equation}\label{Tdual_Lax}
\tilde{\mathcal{L}}_{\pm}:=\frac{\tilde{j}_{\pm}\pm z\tilde{\mathfrak{J}}_{\pm}}{1-z^2}
\end{equation}
has a curvature
\begin{align}
&\partial_{+}\tilde{\mathcal{L}}_{-}-\partial_{-}\tilde{\mathcal{L}}_{+}+[\tilde{\mathcal{L}}_{+},\tilde{\mathcal{L}}_{-}]
\\
& = \frac{1}{1-z^2}\Bigl( \partial_{+}\tilde{j}_{-}-\partial_{-}\tilde{j}_{+}+[\tilde{j}_{+},\tilde{j}_{-}]-z(\partial_{+} \tilde{\mathfrak{J}}_{-} + \partial_{-} \tilde{\mathfrak{J}}_{+})\Bigr) \, ,
\notag
\end{align}
which vanishes if (\ref{T-dual-EOM-X}) is satisfied and $\tilde{\mathfrak{J}}$ is conserved. This establishes a Lax representation for the T-dual model. 

To argue for integrability -- focusing, for simplicity, on the gauging of H$=$G$_{L}$ such that $X=\Lambda$ -- one can construct a canonical transformation between (\ref{deformed-PCM}) and its T-dual (\ref{T-dual-action}). When $E = 0$, and the auxiliaries have been eliminated, it is well-known \cite{Alvarez:1994wj,Lozano:1995jx} that the PCM and T-dual PCM are related by a canonical transformation with type-1 generating function
\begin{align}\label{generating_function}
    \mathcal{F} [ \Lambda, \phi ] = - \int_{\Sigma} \tr \left( \Lambda j_\sigma \right) \, ,
\end{align}
where $\phi$ denotes a set of local coordinates on the Lie group $G$.
One can show that the same generating function (\ref{generating_function}) induces a symplectomorphism that relates the Hamiltonians of the deformed models (\ref{deformed-PCM}) and (\ref{T-dual-action}). Remarkably, this holds for any interaction function $E$. Since such a symplectomorphism preserves all Poisson brackets, and since the AFSM is known \cite{Ferko:2024ali} to possess an infinite collection of Poisson-commuting conserved charges, there also exist infinitely many charges in involution for the deformed T-dual PCM. This establishes classical integrability of the deformed T-dual models.

\section{Conclusion and Outlook}
\label{secVI}

In this work, we have presented a case study of the interplay between dualities and deformations in quantum field theory, focusing on the connection between T-duality and stress tensor deformations in $2d$ sigma models. This example is especially interesting because the structure and properties of our deformations take almost exactly the same form on both sides of the duality.

Our analysis exploits the fact that stress tensor deformations are \emph{universal} and \emph{intrinsically defined}. Universality means that they exist in any translation-invariant field theory, since every such theory admits a conserved stress tensor. By ``intrinsically defined'' we mean that the deformation is built via an abstract procedure that does not depend on the 
details of the theory being deformed, except through the initial condition; one may always define the Hilbert stress tensor $T_{\alpha \beta}$ by coupling a theory to 
gravity, and then deform by a function of $T_{\alpha \beta}$. This is the fundamental reason why our deformations behave identically for the PCM and its T-dual, as the coupling to gravity works in the same way for both duality frames.

The take-away lesson of our work is that one can learn more about dual descriptions of QFTs by studying intrinsically defined deformations on both sides of the duality. Several future directions promise to build on this insight and further expand our understanding of dualities and deformations. One avenue is to develop our auxiliary field formalism for other sigma models, such as those on symmetric spaces. Another direction is to study the interplay of these deformations with other dualities, such as S-duality for $4d$ gauge theories, which can also be understood as a canonical transformation \cite{Lozano:1995aq}. A third line of inquiry is to apply auxiliary field techniques to deformations of holographic dualities, such as the $\mathrm{AdS}_3 / \mathrm{CFT}_2$ correspondence. More generally, it will be important to extend our analysis to the quantum level.

We believe that further progress in any of these directions will help to illuminate more features of the space of field theories, and may eventually help point the way toward a radical rethinking of what a QFT really is.

\begin{acknowledgments}
\medskip
\noindent\textbf{Acknowledgements}
We are grateful to Peter Bouwknegt, Mattia Cesaro, Johanna Knapp, Yolanda Lozano, Jock McOrist, Silvia Penati, Anayeli Ramirez, Savdeep Sethi, Alessandro Sfondrini, Dmitri Sorokin, and Martin Wolf for fruitful discussions. 
D.\,B. is supported by Thailand NSRF via PMU-B, grant number B13F670063.  
C.\,F. is supported by U.S. Department of Energy grant DE-SC0009999 and funds from the University of California. 
L.\,S. is supported by a postgraduate
scholarship at the University of Queensland. 
G.\,T.-M. has been supported by the Australian Research Council (ARC) Future Fellowship FT180100353, ARC Discovery Project DP240101409, and the Capacity Building Package of the University of Queensland.
\end{acknowledgments}

\appendix

\bibliographystyle{apsrev4-2} 

%

\end{document}